\documentclass[aps,prl,twocolumn,superscriptaddress,showpacs]{revtex4}
\usepackage{graphicx,amsmath,bm}
\usepackage{amssymb}
\begin{document}

\title{Enhanced thermoelectricity at the ultra-thin film limit}

\author{Nguyen Thi Thu Thao}
\affiliation{Nano and Energy Center, VNU University of Science, Vietnam National University, Hanoi}
\author{Dang Tuan Linh}
\affiliation{Nano and Energy Center, VNU University of Science, Vietnam National University, Hanoi}
\author{Bach Huong Giang}
\affiliation{Faculty of Physics, VNU University of Science, Vietnam National University, Hanoi}
\author{Dang Huu Tung}
\affiliation{Nano and Energy Center, VNU University of Science, Vietnam National University, Hanoi}
\author{Nguyen Trung Kien}
\affiliation{Faculty of Physics, VNU University of Science, Vietnam National University, Hanoi}
\author{Pham Thi Hong}
\affiliation{Nano and Energy Center, VNU University of Science, Vietnam National University, Hanoi}
\author{Nguyen Tran Thuat}
\affiliation{Nano and Energy Center, VNU University of Science, Vietnam National University, Hanoi}
\author{Nguyen Viet Tuyen}
\affiliation{Faculty of Physics, VNU University of Science, Vietnam National University, Hanoi}
\author{Nguyen The Toan}
\affiliation{Faculty of Physics, VNU University of Science, Vietnam National University, Hanoi}
\author{Nguyen Quoc Hung}
\email{hungngq@hus.edu.vn}
\affiliation{Nano and Energy Center, VNU University of Science, Vietnam National University, Hanoi}
\affiliation{Nanotechnology Program, Vietnam-Japan University, Vietnam National University, Hanoi}

\begin{abstract}
At the ultra-thin film limit, quantum confinement strongly improves thermoelectric figure of merit in materials such as Sb$_2$Te$_3$ and Bi$_2$Te$_3$. These high quality films have only been realized using well controlled techniques such as molecular beam epitaxy. We report a two fold increase in the Seebeck coefficient for both p-type Sb$_2$Te$_3$ and n-type Bi$_2$Te$_3$ using thermal co-evaporation, an affordable approach. At the thick film limit greater than 100 nm, their Seebeck coefficients are around 100 $\mu V/K$, similar to results obtained in other work. When the films are thinner than 50 nm, the Seebeck coefficient increases to about 500 $\mu V/K$. With a total Seebeck coefficient $\sim$ 1 mV/K and an estimate ZT $\sim$ 2, this pair of materials is the first step to a practical micro-cooler at room temperature. 
\end{abstract}

\maketitle

When electrons move inside a solid, they carry both electric currents and heat currents. The twin mechanisms: Peltier effect and Seebeck effect hold a huge potential for practical applications \cite{Rowe}. Being compatible with nano fabrication techniques, they can be included as an active heat pump to prevent integrated circuit (IC) chip from overheating \cite{Iyengar}. It could be the alternate technology that harvests energy, especially when heat waste from fossil fuel is a dominant factor for global warming \cite{SueTsai}. In the absence of gravity, thermoelectric machines work in deep space without a moving part \cite{Voyager}, where no other cooling technology exists. Despite the demands, thermoelectric materials perform deficiently when compares to other cooling technologies. Using electric current to carries the heat away, they suffer heavily from the back flow of heat. The figure of merit for a thermoelectric material, therefore, involves the electrical conductivity $\sigma$, the thermal conductivity $\kappa$, and the Seebeck coefficient $S=\frac{\Delta V}{\Delta T}$: $ZT=S^2\sigma T/\kappa$, where the temperature term is added such that $ZT$ is a dimensionless number. With all parameters $\sigma,\kappa$, and $S$ depend on each other and on temperature, it is difficult to optimize $ZT$. To date, the world record value for $ZT$ is 2.6 for an epitaxial superlattice made of Sb$_2$Te$_3$ and Bi$_2$Te$_3$ \cite{Venkatasubramanian}. 

Of all candidates, chalcogenides, especially tellurium based, thermoelectric materials are routinely reported with an exceptionally high $ZT$ both in bulk materials and in thin films. Due to a high degeneracy at the edges of the energy band, p-type Sb$_2$Te$_3$ and n-type Bi$_2$Te$_3$ \cite{Rowe} possess an exceeding number of carriers that contribute to the thermoelectric effect and are an excellent pair for cooling application near room temperature. Grown using bulk techniques \cite{HuongJAC,NongRSC,TuanNcom,UrePR}, single crystal ingots achieve a high Seebeck coefficient, up to 500 $\mu V/K$, a high electrical conductance, and a low thermal conductivity. To improve ZT, carrier concentration and mobility can be tuned through doping or optimizing various growth conditions. Despite all the challenges stem from the material engineering, high quality bulk thermoelectric ingots are produced on the industrial scale with commercial products readily available \cite{TEC}.

\begin{table}
	\begin{center}
\begin{tabular}{|c|c|c|c|c|}\hline
	series & $S_{\rm{max}} $ & d & $S_{\rm{saturated}}$ & Note \\
	& $\mu V/K$ & nm & $\mu V/K$ & \\\hline
	 S1& 135 & 24  &  34 & initial observation\\  
	 S2& 502 & 12 & 90 & reproduced, optimized \\  
	 B1& 134 & 28 & N/A & initial observation \\ 
	 B2& 407 & 28 & 130 & reproduced, not systematic \\ 
	 B3& 200 & 28 & 155 & reproduced, systematic study \\ 
	 B4& 401 & 28 & 15 & reproduced, optimized \\ 
	\hline
\end{tabular}
\end{center}
\caption{Different thickness series for Sb$_2$Te$_3$ (S1, S2) and Bi$_2$Te$_3$ (B1 to B4). The highest Seebeck coefficient, its corresponding thickness, and the saturated Seebeck coefficient at the thick film limit are listed. After the first studies where the enhancements in Seebeck are observed in both materials, we perform more systematic experiments where both fabrication and measurement conditions are optimized and strictly controlled. Series B2 focused on the evaporation rate, and therefore not really systematic. Series B3 had an oil leak with the vacuum pump in the measurement setup.}
\end{table}

However, thermoelectricity is not necessary an intrinsic material property. At the nanometer scale, quantum confinement increases $S$, scattering reduces $\kappa$, and a high carrier concentration improves $\sigma$ \cite{Vineis,RenQM16}. All in all, theory \cite{Hicks1D, Hicks2D} predicts that $ZT$ is expected to largely enhanced when the critical size is on the order of a monolayer, smaller than 10 nm. This strategy was quickly implemented in experiments with astonishing results. Various approaches that take advantage of the size effect have been employed such as superlattice growth using molecular beam epitaxy \cite{Venkatasubramanian}, hot press nanopowder \cite{Poudel}, quantum dots \cite{Harman}, nanostructured \cite{Biwas}, or exfoliation \cite{Balandin}. Micron size cooling devices \cite{Snyder} are also reported with outstanding performances. Interestingly, the recent discovery of Sb$_2$Te$_3$ and Bi$_2$Te$_3$ as topological insulators \cite{SCZhangTopo} provides another mechanism for the improvement of thermoelectricity. When the surface to volume ratio increases at the nanometer scale, a hybridization induced band gap appears and edge states dominate transport. In the vicinity with the Fermi level, these states support asymmetric transport of hot and cold carriers, thus an enhancement in $S$ \cite{SCZhangPRL, Ghaemi}. In practice, the growth of these high quality materials requires special expertise with sophisticated apparatuses which is difficult, expensive and not scalable.

\begin{figure}[t]
\begin{center}
\includegraphics[width=3in,keepaspectratio]{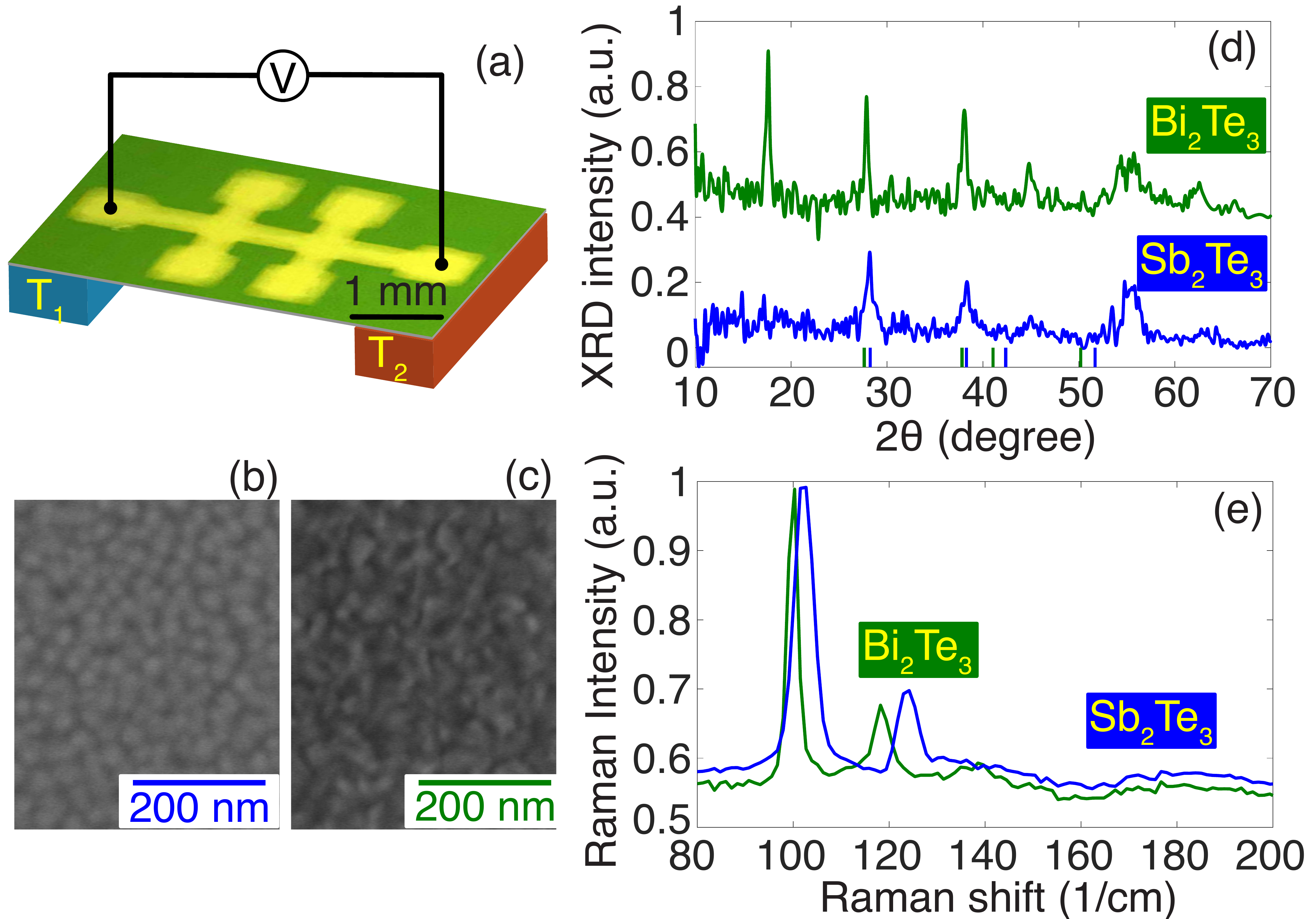}
\caption{(a) Optical micrograph of a film used in the experiment, embedded in a measurement diagram. A temperature gradient $\Delta T=T_2-T_1$ is generated by two commercial thermoelectric coolers. Thermoelectric coefficient is measured in a homemade vacuum chamber with help from a computer and standard electronics. (b) Scanning electron micrograph of the 24 nm Sb$_2$Te$_3$ film and (c) micrograph of the 28 nm Bi$_2$Te$_3$ thin film showing poly-crystal films with grain size about 40 nm. (d) X-ray diffractogram for Sb$_2$Te$_3$ (blue), thickness 48 nm and Bi$_2$Te$_3$ (green), thickness 69 nm. The standard x-ray peaks are marked at the bottom of the x axis. (e) Raman spectroscopic shift of the 48 nm Sb$_2$Te$_3$ films (green), and 69 nm Bi$_2$Te$_3$ films (blue).}
\label{fig1}
\end{center}
\end{figure}

Alternatively, thermoelectric thin films can be made using more accessible methods such as thermal evaporation \cite{Rogacheva}, co-evaporation  \cite{GoncalvesBiTe,GoncalvesSbTe}, sputter \cite{LeeAIP18}, chemical vapor deposition \cite{MOVPE}, or laser pulse deposition \cite{PhuocPLD}. Lack of rigorous controls, these films suffer an inhomogeneous stoichiometry and uncontrolled polycrystal structures. Their thermoelectricity are low-quality and are yet practical. With all parameters optimized, such as  substrate temperature, quality of the initial materials, or post-evaporation processes like annealing, Sb$_2$Te$_3$ and Bi$_2$Te$_3$ thin films can obtain a Seebeck coefficient up to 200 $\mu V/K$ \cite{CaiAPS13}. To reduce the back flow of heat, the film is often made in the range of $\mu m$ \cite{LeeAIP18,GoncalvesBiTe,TaeSung}, and thus are not  benefited from the quantum size effect. Despite predictions from theory \cite{Hicks1D, Hicks2D, SCZhangPRL, Ghaemi, kim2009, WangSciRep,ZahidAPL}, experimental works are rare at the ultra-thin film limit. We are only aware of a single group who measures Seebeck coefficient in this ultra-thin regime. Agree with semiclassical theory \cite{Hicks2D,Freik}, multiple quantum oscillations in transport and thermoelectric properties as a function of film thickness are observed. In these works, various materials, including Bi$_2$Te$_3$ \cite{RogachevaBiTe,RogachevaBiTe2}, PbTe \cite{RogachevaPbTe}, or SnTe \cite{RogachevaSnTe}, are thermally evaporated on glass substrates from a single stoichiometry source. They all show oscillations of the Seebeck coefficient as a function of film thicknesses, but none exhibits a clear increment in the Seebeck coefficient, an important feature that was predicted theoretically \cite{WangSciRep,Hicks1D,Hicks2D,Ghaemi,SCZhangPRL}.

In this paper, we report a strong increase in the Seebeck coefficient as the films approach the ultra-thin limit in two materials: p-type Sb$_2$Te$_3$ and n-type Bi$_2$Te$_3$. Six thickness series of both materials are deposited on silicon substrates using thermal co-evaporation with optimized parameters. They all show a strong increase in the Seebeck coefficient and the power factor $\digamma=S^2\sigma$ as the films get to the ultra-thin limit. At the thick film limit greater than 100 nm, the Seebeck coefficients return to their bulk values for both materials. We verify this observation using Boltzmann calculations \cite{Hicks2D}. Although the two materials Sb$_2$Te$_3$ and Bi$_2$Te$_3$ are different, the similarity in their thermoelectric properties emphasizes the importance of the quantum confinement effect in these topological insulators. We choose to report them together and make the distinction when needed.

\begin{figure}[t]
\begin{center}
\includegraphics[width=3in,keepaspectratio]{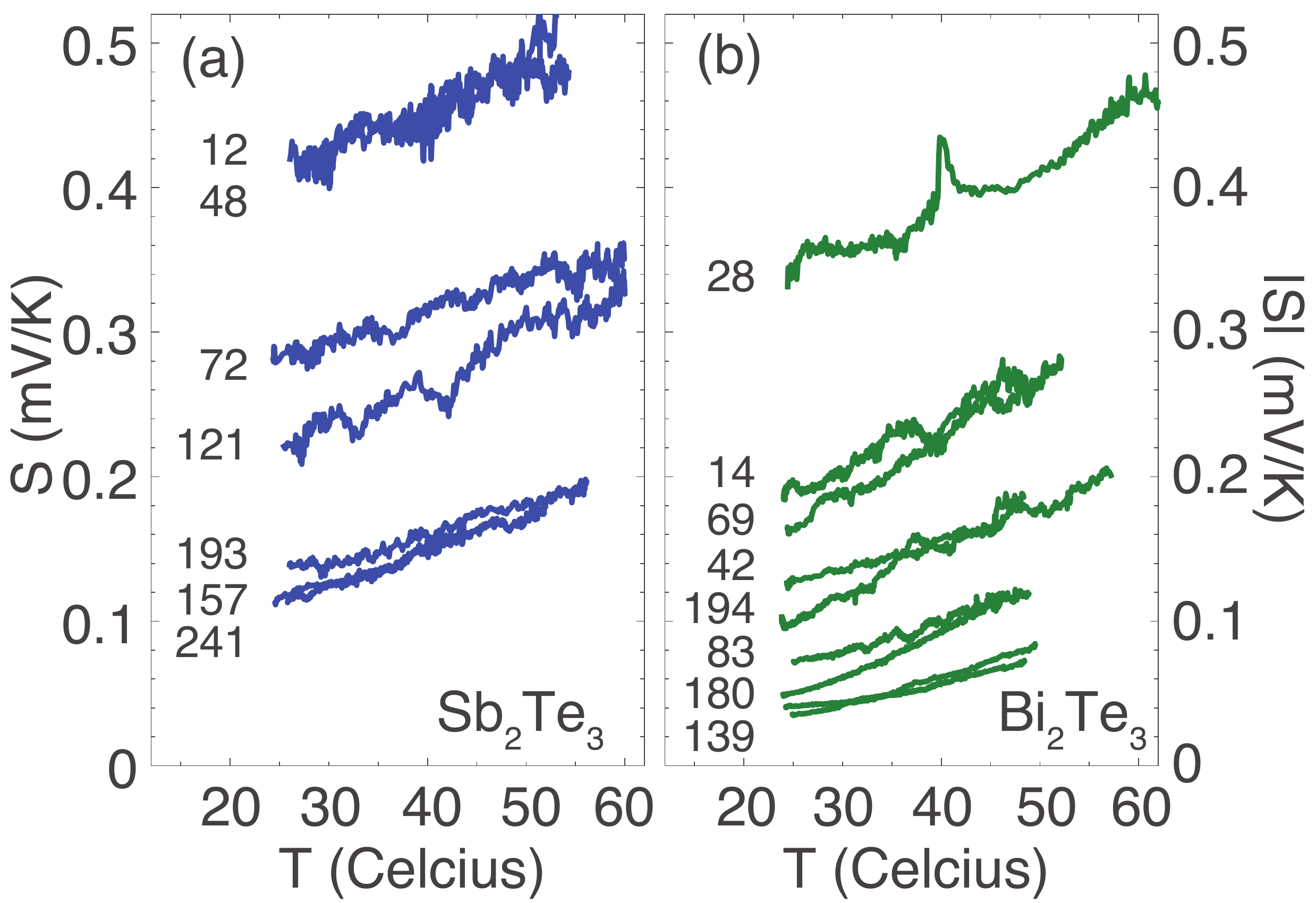}
\caption{(a) Seebeck coefficient for p-type Sb$_2$Te$_3$ series S2 and (b) absolute Seebeck coefficient for n-type Bi$_2$Te$_3$ series B4 as a function of temperature at various thicknesses. $S>0$ for all Sb$_2$Te$_3$ and $S<0$ for all Bi$_2$Te$_3$ films. The numbers next to the curves indicate the film's thicknesses in nm.}
\label{fig2}
\end{center}
\end{figure}

The thin Sb$_2$Te$_3$ (Bi$_2$Te$_3$) films are thermally co-evaporated from two separate sources antimony (bismuth) and tellurium through a metallic shadow mask with a five square Hall bar geometry, see Fig. 1a. All materials are 5N quality. Prior to the result reported in this paper, deposition parameters such as material flow rates, their ratio, and substrate temperatures are carefully optimized \cite{Dang,Thao}. Throughout the evaporation, it is important to maintain the flow ratio of the two materials, Te over Sb (Bi), at precisely 2.5, which is independently monitored using two quartz crystals. To ensure the proper stoichiometry, the Si substrate temperature is fixed at 473 K for Sb$_2$Te$_3$ and 523 K for Bi$_2$Te$_3$ during the evaporation. After deposition, all film thicknesses are confirmed with a profilometer. Basic transport parameters such as concentration and mobility are obtained from a Hall measurement system at 1 T magnetic field at room temperature, see value of some typical films in table II. Prior to all measurements, the 1.5 cm $\times$ 1.5 cm silicon substrate with a thin layer of SiO$_2$ on top are confirmed as an electrical isolators. Seebeck coefficient $S=\Delta V/\Delta T$ is measured in a homemade apparatus where a temperature difference is provided homogeneously by two copper blocks, whose temperatures are precisely controlled using two commercial thermoelectric coolers \cite{TEC}. This measurement is performed inside a metallic vacuum chamber for better electrical signal and thermal isolation, especially from convection. To ensure consistency, a series of films with different thicknesses is fabricated and measured consecutively before the evaporator is used for other purposes. In total, two series of Sb$_2$Te$_3$ and four series of Bi$_2$Te$_3$ are measured, as shown in table I. These result are quite similar and we only report series S2 for Sb$_2$Te$_3$ and series B4 for Bi$_2$Te$_3$ in this paper.

\begin{figure}[t]
\begin{center}
\includegraphics[width=3in,keepaspectratio]{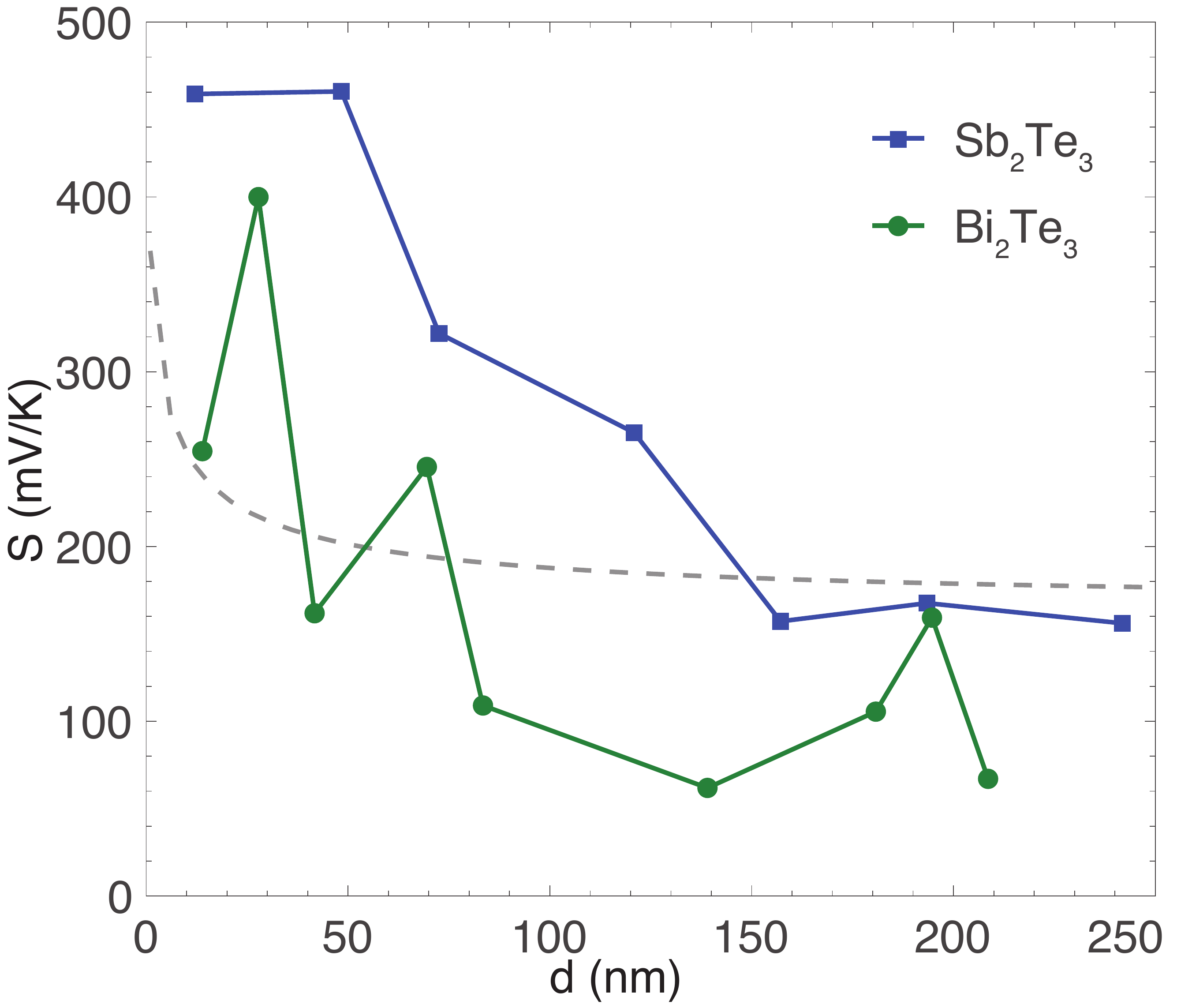}
\caption{Seebeck coefficient for Sb$_2$Te$_3$ (blue squares) and Bi$_2$Te$_3$ (green circles) as a function of film thickness measured at temperature 316 K. All film thicknesses are measured with a profilometer. The dashed line is a theoretical calculation according to Ref. \cite{Hicks2D}.}
\label{fig3}
\end{center}
\end{figure}

All films' structure are analyzed using x-ray diffraction and Raman spectroscopy. Except for the thinnest film near 20 nm, standard XRD peaks near 27 degree are observed for both Sb$_2$Te$_3$ and Bi$_2$Te$_3$, validate the proper stoichiometry of the films. Similarly in Raman spectroscopy with an excitation at 632.8 nm, peaks near 120 cm$^{-1}$ are observed homogeneously over all area of the films. We find a strong correlation between these peaks with thermoelectric effect, and use them to screen the films for thermoelectric measurements. Namely, about a third of our films does not show proper crystalline peaks, even with optimized evaporation parameters, and hence, does not have high Seebeck coefficient. To obtain the grain size, we use Scherrer formula: $\delta=\frac{0.9\lambda}{\beta cos\theta}$, where $\lambda$ = 1.5406 \AA\ is the x-ray wavelength, $\beta$ is the full width at half maximum of the peak, and $\theta$ is the Bragg angle, as in Fig. 1b. From the diffractogram peaks, the grain size is estimated around 40 nm with a weak dependence on film thickness. This value is closed to the grain size as directly observed from the scanning electron micrograph, Fig. 1 b  and c. Both methods show that films of different thicknesses, up to 1 $\mu m$, maintain an almost constant grain size.

\begin{table}
	\begin{center}
\begin{tabular}{|c|c|c|c|c|c|c|}\hline 
	id & d & $\delta$ & $\sigma$ & n$\times 10^{19}$ & $\mu$  & S  \\
	 & nm & nm & ($\Omega$cm)$^{-1}$ & cm$^{-3}$ & cm$^2/Vs$ & $\mu V/K$ \\
	\hline\hline
	Sb$_2$Te$_3$ & 12 & 35.2 & 26.4 &9.5 & 2.2 & 502   \\	
	Sb$_2$Te$_3$ & 24 & 44.3 & 11.8 & 0.3 & 1.1 & 395   \\	
	Sb$_2$Te$_3$ & 241 & 47.7 & 27.5 & 0.5 & 0.5 & 90   \\	
	
	Bi$_2$Te$_3$ & 14 & 38 & 71.9 & -22.3 & 37.1 & -254   \\	
	Bi$_2$Te$_3$ & 28 & 37.5 & 22.5 & -1.4 & 22.7 & -400   \\	
	Bi$_2$Te$_3$ & 180 & 38.8 & 102.5 & -11.4 & 13.4 & -105   \\	
\hline
\end{tabular}
\end{center}
\caption{A list standard parameters for typical films of both materials Sb$_2$Te$_3$ and Bi$_2$Te$_3$. Here, $d$ is film thickness, $\delta$ is the grain diameter, $\sigma$ is the electrical conductivity, $n$ is the carrier concentration, $\mu$ is the mobility, and $S$ is the Seebeck coefficient. $\delta$ is measured with a contrast threshold algorithm using SEM images and verified with Scherer formula; $n$, $\sigma$, and $\mu$ are measured from 0 to 1 T perpendicular magnetic field at room temperature using a Hall measurement system; and $S$ is measured in our homemade apparatus.}
\end{table}

We find that the Seebeck coefficient strongly increases as the film thickness approaches the ultra-thin limit in both materials. In Fig. 2, all Sb$_2$Te$_3$ films are p-type with $S > 0$ and all Bi$_2$Te$_3$ films are n-type with $S < 0$. For simplicity, we use the absolute value $|S|$ for Bi$_2$Te$_3$. Near room temperature, the Seebeck coefficients slightly increase with temperature for both Sb$_2$Te$_3$ and Bi$_2$Te$_3$, similar to other materials of the same family \cite{NongRSC,PhuocPLD, Rowe, TuanNcom,Biwas}. For films with thicknesses greater than 100 nm, the Seebeck coefficients $|S|$ is less than 200 $\mu V/K$, agree with results obtained in thin films made from other approaches \cite{NongRSC,LeeAIP18,GoncalvesSbTe,GoncalvesBiTe,TaeSung,Rogacheva,PhuocPLD}. However, when the films are thinner than 100 nm, the Seebeck coefficients increase and reach 500 $\mu V/K$, a value only obtained in single crystal bulk \cite{HuongJAC}. We emphasize this result in Fig. 3, where the Seebeck coefficient at temperature 316 K is plotted against film thickness. It should be noted that this strong dependence of $|S|$ on film thickness is observed in all 6 series of films for both materials, underlines the significance of quantum confinement \cite{Hicks1D,Hicks2D} and the topological band gap \cite{Ghaemi,SCZhangPRL} in these tellurium based materials. To verify this observation, our calculation following Ref. \cite{Hicks2D} is showed as dashed lines in Fig. 3. 

\begin{figure}[t]
\begin{center}
\includegraphics[width=3in,keepaspectratio]{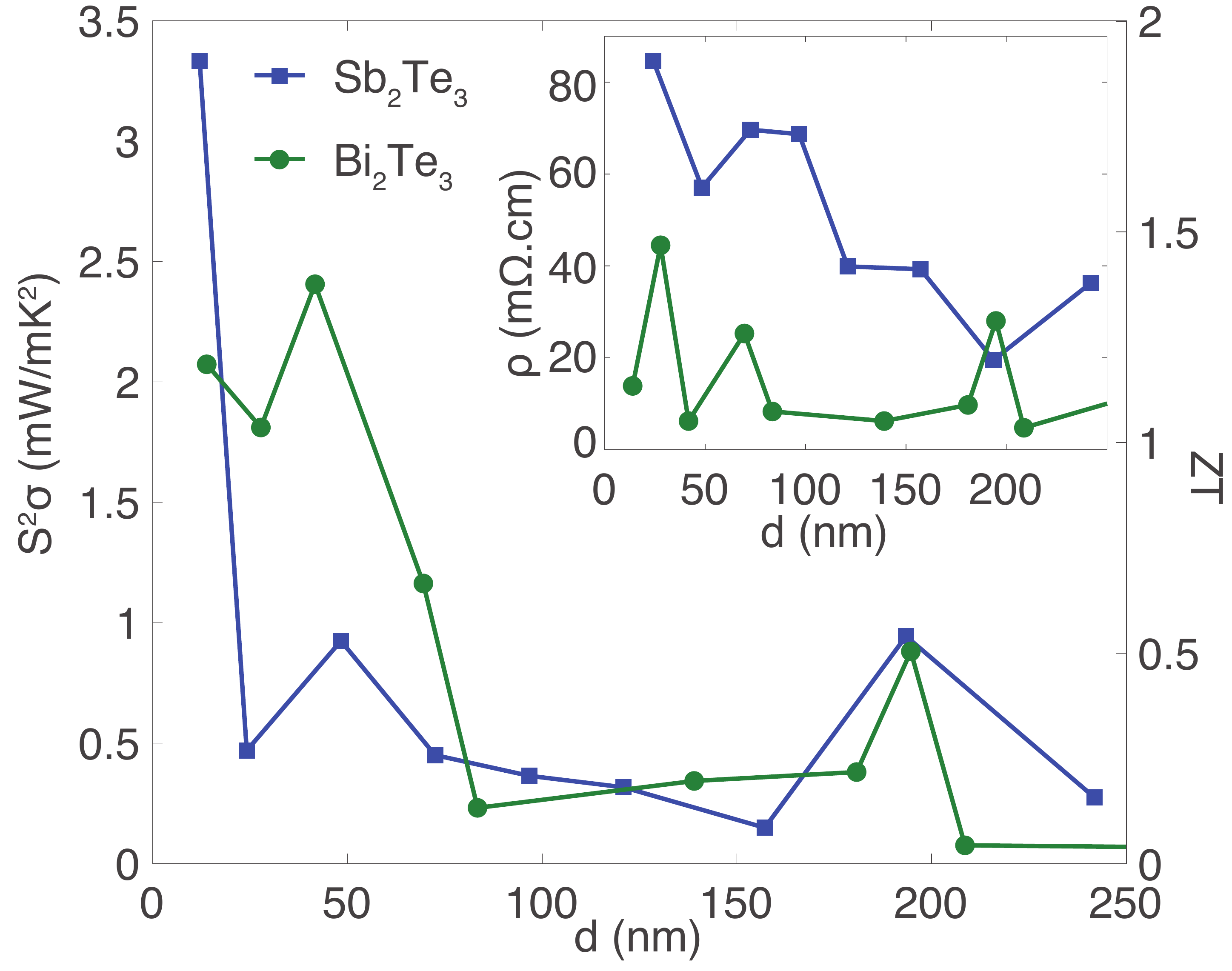}
\caption{Power factor $S^2\sigma$ as a function of film thickness for Sb$_2$Te$_3$ (blue squares) and Bi$_2$Te$_3$ (green circles). We estimate $ZT$ assumed that $ k=0.5$ W/mK \cite{RuanAPL,WangSciRep} for all films. In the inset, resistivity as a function of film thickness is shown for both materials.}
\label{fig4}
\end{center}
\end{figure}

It is natural that resistivity $\rho$ increases as the film becomes thinner, as showed in the inset of Fig. 4. However, the strength of a thermoelectric material lies in its power factor $\digamma=S^2\sigma$. In Fig.4, $\digamma$ for both Sb$_2$Te$_3$ and Bi$_2$Te$_3$ increase strongly toward the ultra-thin film thickness, showing a similar trend to the Seebeck coefficient in Fig. 3. On the right axis, we assume a constant thermal conductance $\kappa$ = 0.5 W/mK \cite{RuanAPL,WangSciRep} for all films, and therefore, obtain the same increase in $ZT=S^2\sigma T/\kappa$ at the ultra-thin film limit. Although this value for $\kappa$ is an optimistic estimate, we anticipate $\kappa$ could be even smaller at these ultra-thin thicknesses. With a total Seebeck coefficient reaching 1 mV/K and a high electrical conductivity, this pair of a p-type and a n-type semiconductor is expected to be suitable for practical applications.

Thermoelectric properties in solid is often derived using two standard approaches: Landauer diffusive formalism \cite{ZahidAPL,SCZhangPRL,Ghaemi} and Boltzmann transport equation \cite{Hicks2D,WangSciRep,Freik}, which are related in the relaxation time approximation \cite{Lundstrom}. Here, we employ the Boltzmann transport equation following the seminal work in Ref. \cite{Hicks2D}. The Seebeck coefficient for a two dimensional gas is given by $S=-\frac{k_B}{e}\large(\frac{2F_{1}}{F_{0}}-\mu^*\large)$ with $F_i=\int_0^{\infty}\frac{\epsilon^id\epsilon}{exp(\epsilon-\mu^*)+1}$ and $\mu^*=\large(\mu-\frac{\hbar^2\pi^2}{2m_zd^2}\large)/k_BT$. Here, $k_B$ is the Boltzmann constant, $m_z$ is the effective mass in the $z$ direction, $\mu$ is the chemical potential, and $d$ is the film thickness. Using a similar set of parameter as in Ref.\cite{Hicks2D}, carrier mobility $\mu=1200$ cm$^2$/Vs, and effective mass $m_x=0.02 m_0$, $m_y=0.4 m_0$, and $m_z=0.01 m_0$, we obtain a theoretical line that closely resembles our experiment data.

Transports are unmeasurable in films that are thinner than 10 nm. It is possible that the effective thickness that actually contributes to transport is thinner than 10 nm, and also these films are below the conductance's percolation threshold. Both Sb$_2$Te$_3$ and Bi$_2$Te$_3$ crystals have a rhombohedral unit cell and a layered structure with five atomic layers, referred as quintuple layers.  Within the range of a few quintuple layers, there are two mechanisms that increase the Seebeck coefficient significantly. First, in the quantized direction $z$, the energy levels strongly depend on film thickness $d_z$ as $\epsilon_z=\frac{\hbar^2\pi^2}{2m_zd_z^2}$. When the film is thin enough, electrons mostly stay in the first sub-band. The total Seebeck coefficient $S=\frac{\sum{\sigma_i S_i}}{\sum{\sigma_i}}$ for a multi-subband system, therefore, is maximized from a single band contribution \cite{Hicks2D}. Second, the topology nature of the bulk electron wave functions in these materials protect the metallic surface states. When the surface to volume ratio increases, the surface states from top and bottom surfaces hybridize and induce a band gap that increase thermoelectric performance \cite{Ghaemi,WangSciRep,SCZhangPRL}.

Our experiments differ in a number of aspects when compare to the series of work by Rogacheva et al. \cite{RogachevaBiTe,RogachevaBiTe2,RogachevaSnTe,RogachevaPbTe}. First, we co-evaporate from two separate sources and can tune the flow ratio, thus film stoichiometry, better. This is important, as thermal evaporation doesn't preserve material stoichiometry nor ensure film uniformity, especially for two elements with different vapor pressure \cite{GoncalvesBiTe,GoncalvesSbTe}. Second, we use silicon substrate with a top layer of SiO$_2$ instead of glass substrate \cite{RogachevaBiTe2}. The different dangling bonds provided by the SiO$_2$ might affect the quality of each single crystal, especially when the film is close to the percolation threshold. These experimental details are the sources to the differences in the film transport properties, including thermoelectricity. Unlike Rogacheva \cite{RogachevaBiTe,RogachevaSnTe}, we do not observe any oscillation in S as a function of film thickness. Our data points are too scarce to observe the effect. The wiggled in Fig. 3 could as well come from noise or other artifact.

Thermal evaporation is popular for its versatility and flexibility, but not for the precision and quality. We make sure that the measurement data is accurate to the best of our capability. At the starting of a new experiment, multiple details were changed randomly, including upgrading the measurement apparatus or using a different batch of materials and Si substrate. These randomized factors minimize any systematic error we might have, and ensure the reproducibility of the result. It is important to note that the yield is not very high. Even with an identical recipe, about a third of the film doesn't possess proper crystalline peaks in XRD, and have almost no thermoelectric effect. Thermoelectricity is always cross-checked with material analysis. Both XRD and SEM analysis prove that stoichiometry and grain size do not change as a function of film thickness. The fact that we observe a similar effect on two different materials, Sb$_2$Te$_3$ and Bi$_2$Te$_3$, confirms the size effect in thermoelectricity, agrees with theoretical proof from other work and our own. 

In summary, we report a low-cost method that produces thin film of exceptionally high thermoelectricity. Using thermal co-evaporation, the Seebeck coefficients in both p-type Sb$_2$Te$_3$ and n-type Bi$_2$Te$_3$ increase more than two fold with a total S reaches 1 mV/K at the ultra-thin film limit. Agree with theoretical predictions, this enhancement is a consequence of the quantum confinement effect in the topological insulator materials.

We acknowledge the support from the National Foundation for Science and Technology Development through Grant Number 103.02-2015.79. Samples were fabricated and measured at the Nano and Energy Center, VNU University of Science, Vietnam National University, Hanoi.


\begin{thebibliography}{70}
\bibitem{TEC} Thermoelectric coolers are available at a price around $\$10$ per piece on popular commercial platforms such as Amazon or Alibaba.
\end{thebibliography}
\end{document}